\documentclass[reprint,amsmath,amssymb,aps,prl,superscriptaddress,floatfix,nobalancelastpage,raggedbottom]{revtex4-2}
\usepackage{graphicx}
\usepackage{dcolumn}
\usepackage{bm}
\usepackage{tipa}
\usepackage{bbm}
\usepackage{latexsym}
\usepackage{physics}
\usepackage{booktabs}
\usepackage[percent]{overpic}
\usepackage{newtxmath}
\usepackage{caption}
\usepackage{subcaption}

\usepackage{xcolor}

\definecolor{apsblue}{RGB}{0,0,150}

\usepackage[
  colorlinks=true,
  allcolors=apsblue,
  pdfborder={0 0 0}
]{hyperref}

\makeatletter
\newcommand{\EndMatterTitle}{%
  \par\addvspace{1.5ex}%
  \onecolumngrid@push
  \begingroup
    \centering\large\bfseries End Matter\par
  \endgroup
  \nobreak\@nobreaktrue
  \addvspace{1.5ex}%
  \onecolumngrid@pop
}
\makeatother

\newcommand{\beginendmatter}{%
  \EndMatterTitle
  \setcounter{equation}{0}%
  \renewcommand{\theequation}{A\arabic{equation}}%
}

\usepackage[doipre={doi:~}]{uri}

\begin{document}

\title{Charge-Conjugation Violation and Population Asymmetry in Bipartite Fermionic Lattices}
\author{Di Xiao}
\affiliation{National Gravitation Laboratory, MOE Key Laboratory of Fundamental Physical Quantities Measurement, and School of Physics, Huazhong University of Science and Technology, Wuhan 430074, People's Republic of China}
\author{Xue-Ting Fang}
\affiliation{National Gravitation Laboratory, MOE Key Laboratory of Fundamental Physical Quantities Measurement, and School of Physics, Huazhong University of Science and Technology, Wuhan 430074, People's Republic of China}
\author{Lushuai Cao}  \email[E-mail: ]{lushuai\_cao@hust.edu.cn}
\affiliation{National Gravitation Laboratory, MOE Key Laboratory of Fundamental Physical Quantities Measurement, and School of Physics, Huazhong University of Science and Technology, Wuhan 430074, People's Republic of China}
\author{Zhong-Kun Hu}  \email[E-mail: ]{zkhu@mail.hust.edu.cn}
\affiliation{National Gravitation Laboratory, MOE Key Laboratory of Fundamental Physical Quantities Measurement, and School of Physics, Huazhong University of Science and Technology, Wuhan 430074, People's Republic of China}
\affiliation{Wuhan Institute of Quantum Technology, Wuhan 430206, People’s Republic of China}
\author{Peter Schmelcher}  \email[E-mail: ]{peter.schmelcher@uni-hamburg.de}
\affiliation{Center for Optical Quantum Technologies, Department of Physics, University of Hamburg, Luruper Chaussee 149, 22761 Hamburg Germany}

\begin{abstract}
Charge conjugation violation (CCV) is a fundamental concept in particle physics 
and has been observed in quasiparticle excitations of quantum many-body systems,
which typically relies on an embedded external symmetry breaking to the underlying system. 
An open question is how an intrinsic CCV mechanism could emerge and 
what its macroscopic consequences would be. We establish sublattice kinks in bipartite fermionic
lattices as a concrete setup showing intrinsic CCV.
The intrinsic CCV of the sublattice kink arises from the graph-topological
nature of the underlying Hamiltonian, with no explicit symmetry breaking taking place.
It leads to a population asymmetry of different configurations and imprints 
a hidden leaf-like structure in the eigenenergy spectrum.
The population asymmetry also leads to an imbalanced sublattice-kink production triggered by
the vacuum-instability in the quench dynamics.
Our work demonstrates the graph-topology nature of the underlying Hamiltonian
as the microscopic origin of intrinsic CCV, with the population asymmetry as the macroscopic consequence, 
of which the proposed setup is highly amenable to experimental 
implementation via cold-atom quantum simulators.

\end{abstract}


\maketitle
Kinks, also termed domain walls, are topological defects acting as interfaces between distinct domains 
in various quantum systems \cite{kinks_Ryder.1996, kinks_Manton.2004, kinks_Vachaspati.2006, kinks_PhysRevLett.80.5032, kinks_JMathPhys.7.1218, kinks_PhysRevLett.67.1177, kinks_PhysRevLett.75.930}, and could be considered as a quasiparticle family. 
They can emerge in various quantum systems of low-energy scale, 
such as the $\phi^4$ \cite{kinks_phi-4_PhysRevA.46.5214, kinks_phi-4_EurPhysJC.84.1266}
and sine-Gordon models \cite{kinks_SG_PhysRevA.45.6019, kinks_SG_PhysRevB.106.075102} 
as well as spin chains \cite{kinks_spin_NatPhys.13.246, kinks_spin_NatCommun.13.7663, kinks_material_Science.327.177, kinks_material_PhysRevLett.112.137403, kinks_material_PhysRevB.83.020407, kinks_material_JStatMech:TheoryExp.2010.P07015, kinks_spin_atom_NatPhys.20.558, kinks_spin_atom_PhysRevB.102.014308, kinks_spin_atom_PhysRevX.12.031037, kinks_spin_ion_PhysRevLett.122.150601, kinks_spin_ion_NatPhys.17.742, kinks_spin_ion_PhysRevX.13.031017, kinks_spin_sc_Science.325.722, kinks_spin_sc_PhysRevX.5.021027, kinks_spin_sc_PhysRevLett.112.200501, kinks_spin_sc_PhysRevA.82.052311,kinks_spin_atom_EurophysLett.110.26004, kinks_spin_atom_PhysRevA.105.053308, kinks_spin_atom_PhysRevA.110.033316}, 
which can be implemented in condensed matter \cite{kinks_material_Science.327.177, kinks_material_PhysRevLett.112.137403, kinks_material_PhysRevB.83.020407, kinks_material_JStatMech:TheoryExp.2010.P07015} and ultracold atomic settings \cite{kinks_spin_atom_NatPhys.20.558, kinks_spin_atom_PhysRevB.102.014308, kinks_spin_atom_PhysRevX.12.031037, kinks_spin_atom_EurophysLett.110.26004, kinks_spin_atom_PhysRevA.105.053308, kinks_spin_atom_PhysRevA.110.033316}.
In low-energy quantum systems they have been explored as a testbed for various fundamental phenomena \cite{QS_RevModPhys.86.153, QS_PRXQuantum.4.027001, QS_NatRevPhys.5.420}.
For instance, magnetic kinks in spin chains have simulated the confinement-deconfinement transitions \cite{kinks_confinement_PhysRevB.99.195108, kinks_confinement_PhysRevB.102.041118, kinks_confinement_PhysRevB.99.180302, kinks_confinement_NatPhys.21.155, kinks_confinement_PRXQuantum.3.040317, kinks_confinement_EurophysLett.121.37001} 
and collision effects between quarks, as well
as related phenomena, such as meson formation \cite{kinks_hadron_scattering_NewJPhys.23.062001, kinks_hadron_scattering_PhysRevRes.4.L032001, kinks_hadron_scattering_PRXQuantum.3.040309, kinks_hadron_scattering_PhysRevA.109.032613} and thermodynamic properties \cite{kinks_hadron_scattering_PhysRevB.104.L201106, kinks_hadron_scattering_PRXQuantum.3.020316}.

Kinks possess naturally opposite configurations, known as kink and antikink, 
which allows the mimicking and test of particle-antiparticle related phenomena, 
such as charge-conjugation violation (CCV) \cite{CCV_Peskin.2018, CCV_PhysRevD.110.030001, CCV_origin_RevModPhys.76.1, CCV_origin_NewJPhys.14.095012, CCV_origin_RevModPhys.80.577}.
The origin of CCV is an open question with different hypotheses, among which one aim is to generate the violation from a symmetric underlying system, 
instead of introducing any explicit symmetry breaking channels. 
While conceptually profound, this faces the challenge that
it is hard to provide a concrete mechanism to generate the CCV from symmetric underpinnings.
Concerning the macroscopic consequences, it is believed that CCV of fundamental particles
is responsible for the particle-antiparticle imbalance, and constitutes a key ingredient to
the Sakharov conditions \cite{CCV_Sakharov} in explaining the matter-antimatter asymmetry
of the universe.

The CCV and associated particle-antiparticle asymmetry has been presented 
for various types of kinks, and the quench-induced kink-antikink imbalanced excitation dynamics,
for instance, has been explored for characterizing the confinement-deconfinement transition of
magnetic kinks \cite{kinks_spin_NatPhys.13.246, kinks_spin_ion_PhysRevLett.122.150601, kinks_spin_ion_NatPhys.17.742}.
However, the CCV normally relies on introducing external symmetry-breaking fields, e.g.
the magnetic field for the magnetic kinks, which explicitly breaks the spin-rotation symmetry.
It remains then an open question whether an intrinsic CCV of symmetry-breaking-free origin
can arise in the kink family and would provide support to the emergent paradigm of CCV.

We propose that in a bipartite fermionic lattice there exist quasiparticles as sublattice kinks,
which offer intrinsic CCV between the kink and antikink configurations. 
The CCV of the sublattice kink is manifested as the difference in the chemical potential between 
the kink and antikink configurations, and arises from the topology nature
of the underlying bipartite fermionic lattice.
The intrinsic CCV provides a concrete mechanism for the emergent paradigm of quasiparticle CCV.
The kink-antikink asymmetry, as a consequence of CCV, leaves its footprints in both the eigenenergy spectrum and
quench dynamics. The kink-antikink asymmetry generates a hidden leaf-like structure in the 
eigenenergy spectrum of the system. Particularly, the Sakharov conditions for the sublattice kink naturally
emerge, and indeed lead to the imbalanced excitation between the kink and antikink configurations.
The bipartite fermionic lattice then enriches the kink family with the sublattice kink of intrinsic CCV,
and also offers an experimentally feasible testbed of both the microscopic origin and the macroscopic
consequence of CCV employing ultracold atomic ensembles.

\begin{figure}
    \centering
    \includegraphics[width=0.95\linewidth]{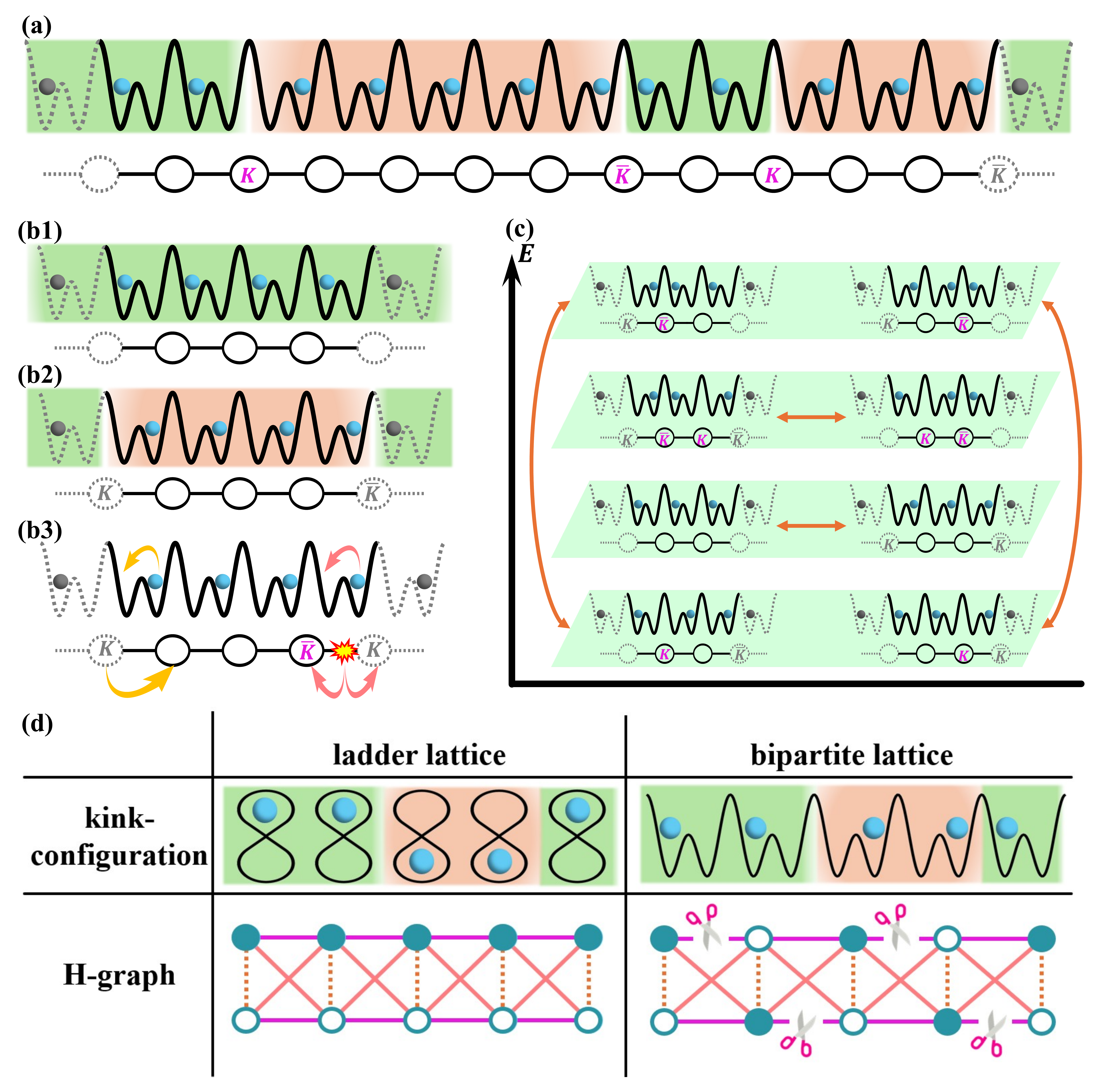}
    \caption{Schematics of SKs with the CCV and kink-number violation.
    (a) the transformation between the bipartite fermionic lattice (upper panel) 
    and the SK bond (lower panel).
    (b1,b2) differentiate between the SK-vacuum states $\ket{\text{vac:L}}_\text{kink}$ (b1)
    and $\ket{\text{vac:R}}_\text{kink}$ (b2) by the virtual SK, and (b3) the virtual-real bond coupling.
    (c) The CCV in terms of the $\text{K}-\bar{\text{K}}$ chemical potential asymmetry shown in 
    both the underlying bipartite lattice and the SK picture.
    (d) The first and second rows illustrate a representative kink configuration and the H-graph,
    respectively, for the CSO ladder lattice (left) and the bipartite lattice (right).}
    \label{fig:1}
\end{figure}

\textit{Sublattice kink and CCV.} We propose and investigate sublattice kinks (SKs) 
with intrinsic CCV in a minimal model of the bipartite fermionic lattice, 
which is composed of spin polarized fermionic particles confined in a one-dimensional 
bipartite lattice with open boundaries. 
We consider the filling of a single fermion per cell, 
and each unit cell of the bipartite lattice contains two sites, 
namely the left (L) and right (R) sites, and all the L(R)-sites form the L(R)-sublattices of the system. 
The fermions are subject to a strong repulsive non-local interaction, 
which energetically favors no double occupancy of each cell. 
Consequently, the low-lying eigenstates and the associated dynamics are effectively confined
to the cell-singly-occupied (CSO) Hilbert subspace, 
which is spanned by basis states with no double occupancy in any unit cell.
The bipartite fermionic lattice of $N$ unit cells, under the CSO truncation,
is governed by the following Fermi-Hubbard Hamiltonian:
\begin{align}
\hat{H}_\text{FH} =& -J \sum_{i=1}^{N} \left( \hat{a}_{i,\text{L}}^{\dagger} \hat{a}_{i,\text{R}} + \text{H.c.} \right) 
- J' \sum_{i=1}^{N-1} \left( \hat{a}_{i,\text{R}}^{\dagger} \hat{a}_{i+1,\text{L}} + \text{H.c.} \right) \nonumber\\
&
+ \sum_{i=1}^{N-1} \sum_{\sigma,\sigma'}V_{\sigma\sigma'}^\text{NNN}\hat n_{i,\sigma}\hat n_{i+1,\sigma'}.
\end{align}
Here, the fermionic operator $\hat{a}_{i,\sigma}^{(\dagger)}$ annihilates (creates) a fermion on 
the $\sigma$-site ($\sigma \in \{\text{L}, \text{R}\}$) of the $i$-th cell, 
and $\hat{n}_{i,\sigma} \equiv \hat{a}_{i,\sigma}^{\dagger}\hat{a}_{i,\sigma}$ is the number operator. 
The first and second terms represent the intra- and inter-cell tunneling with strengths $J$ and $J'$, 
respectively. The last term describes the repulsive non-local interaction, and our discussion
focuses on the truncation of the interaction range to the next-nearest-neighbor (NNN) sites. 
This NNN truncation is introduced to avoid mixing the CCV with other effects,
e.g. the non-local interaction induced kink confinement 
\cite{kinks_spin_NatPhys.13.246, kinks_spin_ion_PhysRevLett.122.150601, kinks_spin_ion_NatPhys.17.742}, 
and we remark that
the results discussed here do not depend on the particular form of the non-local interaction. 

The CSO basis states under the unit filling per cell can be defined as
$|\sigma_1 \cdots \sigma_N\rangle \equiv \prod_{i=1}^{N} \hat{a}_{i,\sigma_i}^{\dagger}
|\text{vac}\rangle_{\text{atom}}$, where $\sigma_i \in \{\text{L}, \text{R}\}$ and $|\text{vac}\rangle_{\text{atom}}$
refers to the vacuum state with no fermions loaded to the bipartite lattice. 
These CSO basis states can be partitioned into sublattice-occupation domains, 
and each domain is composed of atoms occupying the same sublattice sites (L or R) of their respective unit cells, 
and we term a domain where all atoms occupy the left (right) sites as an L-domain (R-domain),
as illustrated by the green (orange) shaded cells in the top panel of Fig. \ref{fig:1}(a).

The boundary between two adjacent domains of opposite types constitutes an SK
as depicted in the lower panel of Fig. \ref{fig:1}(a), of which SKs can be viewed
as a quasiparticle residing on the bonds linking two adjacent cells.
There exist two types of SK configurations, and we define a kink, denoted as $\text K$,
as the interface with an L- and R-domain on its left and right side, respectively.
The antikink ($\bar{\text K}$) then refers to the opposite domain interface configuration
to $\text K$.
$\text{K}$ and $\bar{\text{K}}$ form a particle-antiparticle pair, 
of which the charge-conjugation
operation corresponds to simultaneously switching the sublattice occupation of fermions in all
cells of the underlying bipartite lattice (more detailed description can be found in the End Matter).
Besides, there exist two degenerate SK-vacuum states, as illustrated in Fig. \ref{fig:1}(b1,b2),
which are effectively a single L- and R-domain configuration with no SK excitation, 
and labeled as $\ket{\text{vac:L}}_\text{kink}$ and $\ket{\text{vac:R}}_\text{kink}$, respectively. 

A complete description of SKs can be obtained by transferring $\hat{H}_\text{FH}$ to 
the kink picture, expressed by the creation and annihilation operators of SKs.
For this purpose, we introduce virtual cells to the bipartite lattice 
\cite{superlattice_PhysRevLett.135.101902}, associated with
virtual bonds to SK (c.f. Fig. \ref{fig:1}(a) and the End Matter), to differentiate the degenerate SK-vacuum states. 
As shown in Fig. \ref{fig:1}(b1, b2), $\ket{\text{vac:L}}_\text{kink}$ 
becomes the single vacuum for SK in the real-virtual extended bonds,
and is denoted as $|\widetilde{\text{vac}}\rangle$. 
In the real-virtual extended lattice, $\hat{H}_\text{FH}$ can be mapped to 
$\hat{H}_{\text{kink}}^{\text{ext}} = \hat{H}_\text{r} + \hat{H}_\text{r-v}$, of which
$\hat{H}_\text{r}$ acts on the real bonds and $\hat{H}_\text{r-v}$ corresponds to the coupling between real and virtual bonds:
\begin{subequations}
\begin{align}
    \hat{H}_\text{r} =& \sum_{j=1}^{N-1} \left( \mu_\text{K} \hat{n}_{\text{K},j} + \mu_{\bar{\text{K}}} \hat{n}_{\bar{\text{K}},j} \right) 
-J \sum_{j=1}^{N-2}\left(\hat{\phi}_{j}^{\dagger}+\hat{\bar\phi}_{j}\right)
                   \left(\hat{\bar\phi}_{j+1}^{\dagger}+ \hat{\phi}_{j+1}\right) \\
\hat{H}_\text{r-v} &= \sum_{\alpha\in\{\text{v}_\text{L},\text{v}_\text{R}\}} -J\left(\hat{\phi}_{\alpha}^{\dagger}+\hat{\bar\phi}_{\alpha}\right)
                   \left(\hat{\bar\phi}_{\mathcal{N}_\alpha}^{\dagger}+ \hat{\phi}_{\mathcal{N}_\alpha}\right).
\end{align}
\end{subequations}
In the above equations, $\hat{\phi}_{j}^{(\dagger)}$ and $\hat{\bar\phi}_{j}^{(\dagger)}$ denote the 
annihilation (creation) operators of $\text{K}$ and $\bar{\text{K}}$ on bond $j$ in the extended lattice, respectively.
$\hat{n}_{\alpha,j}=\mathrm{sgn}(\alpha)\hat{\phi}_{j}^{\dagger}\hat{\phi}_{j}+(1-\mathrm{sgn}(\alpha))\hat{\bar\phi}_j^\dagger\hat{\bar\phi}_j$ is the number operator for an 
$\alpha$-type kink on bond $j$, with $\mathrm{sgn}(\alpha)=1$ (0) denoting $\text{K}$ ($\bar{\text{K}}$). 
In $\hat{H}_\text{R}$, the first term refers to the chemical potential of the sublattice kinks,
and the second term describes both the single SK hopping and the $\text{K}-\bar{\text{K}}$ pair excitation/annihilation.
The real-virtual coupling $\hat{H}_\text{r-v}$ includes the hopping and the pair creation/annihilation of sublattice kinks 
between a virtual and its adjacent real bond, 
with $\mathcal{N}_{\text{v}_\text{L}(\text{v}_\text{R})}=1(N-1)$ denoting the neighbor bond of the corresponding virtual bond.

$\text{K}$ and $\bar{\text{K}}$ are antiparticles of each other, with different chemical potentials 
$\mu_{\bar{\text{K}}}=V_\text{RL}^\text{NNN}$ and $\mu_\text{K}=0$,
and this explicitly witnesses the intrinsic CCV between $\text{K}$ and $\bar{\text{K}}$.
The CCV of SK is not induced by any explicit symmetry breaking of $\hat{H}_\text{FH}$,
but the interplay between the geometry of the bipartite lattice and the NNN interaction
of fermions. As illustrated in Fig. \ref{fig:1}(c), 
the lattice geometry assigns $\text{K}$ and $\bar{\text{K}}$ with different 
distances between fermions across the corresponding domain interface, 
and the NNN interaction then leads to different interaction energies of the two spatial
configurations, leading to an intrinsic CCV.
The interplay between the lattice geometry and NNN interactions can be described by
the Hamiltonian graph (H-graph) of $\hat{H}_\text{FH}$, as shown in Fig. \ref{fig:1}(d). 
The H-graph reveals how the lattice geometry affects both the single-particle hopping
and the two-body interactions, and its connectivity imprints the graph
topology to the underlying system.

Figure \ref{fig:1}(d) demonstrates the H-graph topology as the origin of the intrinsic CCV,
by comparing the underlying H-graphs of SK and the ladder-lattice kink, which arises in 
the CSO ladder lattice \cite{kinks_spin_atom_PhysRevA.110.033316,ladder_graph_PRB.110.165110}.
As sketched in the first row of Fig. \ref{fig:1}(d), the CSO ladder lattice 
can exhibit the same kink configuration as the bipartite fermionic lattice, while
the kinks in ladder lattice preserve the charge conjugation.
The different charge-conjugation properties of the two types of kinks lies in the 
graph topology, i.e. the connectivity of the corresponding underlying H-graphs, 
shown in the bottom row of Fig. \ref{fig:1}(d). In both H-graphs, 
the vertices represent the sublattice-occupations of each cell,  
and the occupation in the upper/lower (left/right) sublattices is represented by the 
solid/hollow vertices for the CSO ladder lattice (bipartite fermonic lattice).
The dotted and solid bonds correspond to the intra‑cell hopping and the NNN interaction, respectively.
Transforming the ladder-lattice H-graph to that of the bipartite lattice requires cutting alternating
solid bonds, which alters the connectivity and results in a different H-graph topology of the 
two underlying systems. More importantly, the intrinsic CCV of SK depends solely
on the bond connectivity of the H-graph, instead of the bond strength,
which confirms that the H-graph topology underlies the emergence of intrinsic CCV.

Besides CCV, SK also provide the kink-number violation on the real bonds,
which refers to the non-conservation of the net kink-number $\delta n=n_{\text{K}}-n_{\bar{\text{K}}}$,
with $n_{\text{K}\left(\bar{\text{K}}\right)}$ referring to the number of $\text{K}\left(\bar{\text{K}}\right)$.
The kink-number violation mimics the hadron number violation, and contributes another essential ingredient
of the SK version of Sakharov conditions. The real-bond kink-number violation is attributed to
$\hat{H}_\text{r-v}$, which contributes two channels to the violation. 
The first channel is activated by the hopping of a SK between the virtual and the real bond, 
adding/eliminating a single $\text{K}$ or $\bar{\text{K}}$ to the real bonds. 
This channel is exemplified on the left end of Fig. \ref{fig:1}(b3), which corresponds 
to  the $\hat{\phi}^\dagger_{1}\hat{\phi}_{\text{v}_\text{L}}$ term in $\hat{H}_\text{r-v}$. 
The right end of Fig. \ref{fig:1}(b3) corresponds to the process
of $\hat{\phi}_{1}^{\dagger}\hat{\bar\phi}_{\text{v}_\text{L}}^{\dagger}$,
and illustrates the second channel through a $\text{K}-\bar{\text{K}}$ pair production
on a virtual bond and its neighbor real bond, which also changes the net
kink number on the real bonds. 

\begin{figure}[tbh]
    \centering
    \includegraphics[width=0.95\linewidth, trim=0 0 0 0, clip]{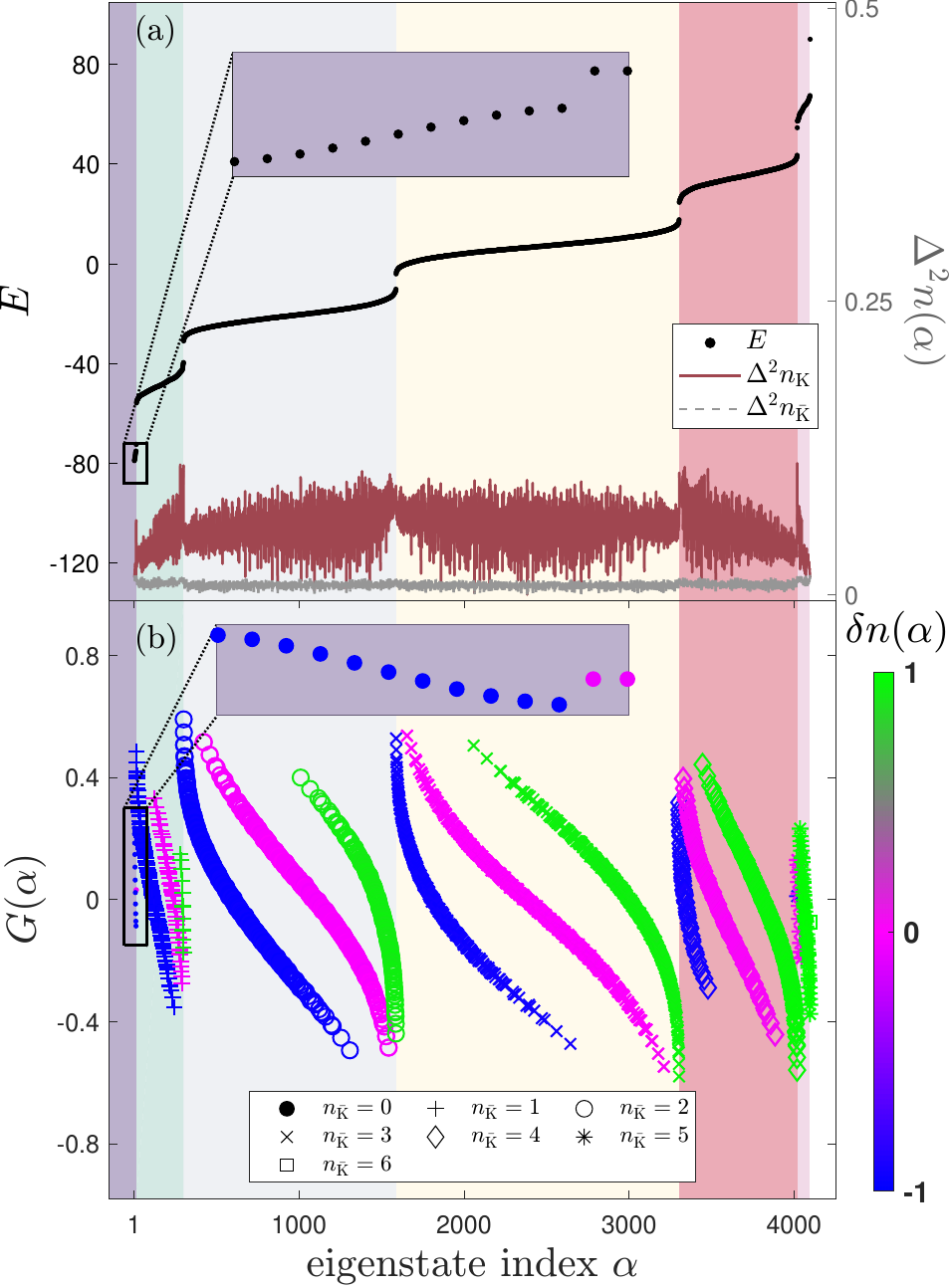}
    \caption{(a) The eigenenergy spectrum (left axis) and the fluctuations of $n_{\bar{\text{K}}}$ (right axis).
    (b) The leaf-like structures shown in light of the intra-cell correlation, 
    with the color and marker type indicating the $\delta n$ and $n_{\bar{\text{K}}}$, respectively,
    of each eigenstate. The colored background of both figures links the band and the leaf associated to
    the same eigenstates. The insets in both figures zoom into the lowest band.}
    \label{fig:2}
\end{figure}

\textit{$\text{K}$-$\bar{\text{K}}$ asymmetry in eigenenergy spectrum.} 
The interplay of the intrinsic CCV and kink-number violation 
of SK leads to the $\text{K}-\bar{\text{K}}$ asymmetry, and imprints a rich structure
in the eigenenergy spectrum of the low-lying eigenstates.
To demonstrate this effect, we numerically calculate the low-lying eigenstates of a bipartite
fermionic lattice of $N=12$ cells with the unit-filling per cell, and the calculation
is done with the original Hamiltonian $\hat{H}_\text{FH}$. 
As displayed in Fig. \ref{fig:2}(a), the calculated low-energy eigenenergy spectrum, at first glance, 
seems rather featureless, apart from being split into multiple bands. However, 
the seemingly featureless eigenenergy spectrum exhibits a hidden structure, 
which becomes visible when examining the single-particle intra-cell correlation $G\left(\alpha\right)$
of a low-lying eigenstate $|\alpha\rangle$, defined as 
$G\left(\alpha\right) \equiv \langle\alpha| \sum_{i=1}^{N}\hat{a}_{i,\text{L}}^{\dagger}\hat{a}_{i,\text{R}}
 + \text{H.c.}|\alpha\rangle/N$. 
$G\left(\alpha\right)$ depicted in Fig. \ref{fig:2}(b) uncovers that each energy band folds up into 
a leaf-like pattern comprising up to three distinct branches, 
and each branch corresponds to a manifold of eigenstates within the band.

Figure \ref{fig:2}(a) (right axis) shows that the variance of $n_{\text K}$ and $n_{\bar{\text K}}$ of
low-lying eigenstates is approaching zero.
This indicates that each low-lying eigenstate is associated with good quantum numbers $\left(n_{\bar{\text{K}}},n_{\text{K}}\right)$
and equivalently $\left(n_{\bar{\text{K}}},\delta n=n_{\text{K}}-n_{\bar{\text{K}}}\right)$. 
In Fig. \ref{fig:2}(b), $n_{\bar{\text{K}}}$ and $\delta n$ are then encoded with the marker type and the
color for the low-lying eigenstates, respectively. 
It can be directly observed that, all eigenstates in a band share the same markers, i.e. the same $n_{\bar{\text{K}}}$,
and those in each branch are of the same $\delta n$, as denoted by the color.
This reveals that the leaf-like structure of each band reflects the explicit $\text{K}-\bar{\text{K}}$ asymmetry
(an analytical result for the low-lying eigenstates and manifold formation is given in the End Matter). 
It is also worth noticing that 
the eigenstates of $\delta n=\pm 1$ in the same band are not energetically degenerate, 
which highlights that the $\text{K}-\bar{\text{K}}$ asymmetry does not result from the spontaneous symmetry breaking, 
but the intrinsic CCV of the sublattice kinks.

The insets of Fig. \ref{fig:2}(a) and (b) zoom into the lowest band of the eigenenergy spectrum, and demonstrate that
the lowest band is split to two subbands. The two eigenstates of the upper subband correspond to
the degenerate $\ket{\text{vac:L}}_\text{kink}$ and $\ket{\text{vac:R}}_\text{kink}$,
and the lower subband is composed of the eigenstates with the SK excitation of a single $\text{K}$.
The higher eigenenergy of $\ket{\text{vac:L}}_\text{kink}$ and $\ket{\text{vac:R}}_\text{kink}$ than that of
the single-$\text{K}$ band suggests the dynamical process mimicking the false-vacuum decay,
in which the $\text{K}-\bar{\text{K}}$ asymmetry can take place.

\textit{Tunable $\text{K}$-$\bar{\text{K}}$ asymmetry in quench dynamics.} 
The false-vacuum-decay like dynamics corresponds to the third part of the complete Sakharov conditions,
i.e. the departure from the thermal equilibrium, to induce the dynamical generation of the particle-antiparticle asymmetry. 
This piece of Sakharov conditions can be fulfilled by the Hamiltonian quench of local tilts applied at the edge sites of the bipartite lattice \cite{addressing_Nature.471.319, addressing_Nature.572.358, addressing_PhysRevLett.114.100503, addressing_Science.352.1562, addressing_PhysRevLett.115.043003, addressing_Nature.621.734, addressing_Science.377.885},
as described by the corresponding Hamiltonian:
\begin{equation}
\hat{H}_{\text{tilt}} = h \left( \hat{n}_{1,\text{L}} + \hat{n}_{N,\text{R}} \right).
\end{equation}
In the above equation, $h$ denotes the strength of the tilt potential. These boundary tilts
lead to a tunable modification of the chemical potentials, given by 
$\mu_\text{K}(h) = \mu_\text{K}\left(h=0\right) + h$ and $\mu_{\bar{\text{K}}}(h) = \mu_{\bar{\text{K}}}\left(h=0\right) - h$.
As a result, the chemical potential imbalance $\delta \mu(h) = \mu_{\bar{\text{K}}}(h) - \mu_\text{K}(h)$ becomes 
continuously adjustable via $h$, as well as the shape of the leaf-like structure of $G(\alpha)$ (details can be found in the End Matter). 
The resonant conditions of $\mu_\text{K} = 0$, $\delta\mu = 0$ and $\mu_{\bar{\text{K}}} = 0$ lead to the resonant coupling between eigenstates of different $(n_\text{K}, n_{\bar{\text{K}}})$, and induce quench dynamics with controlled $\text{K}$-$\bar{\text{K}}$ asymmetry.

\begin{figure}[t]
    \centering
    \includegraphics[width=0.95\linewidth, trim=40 20 10 20]{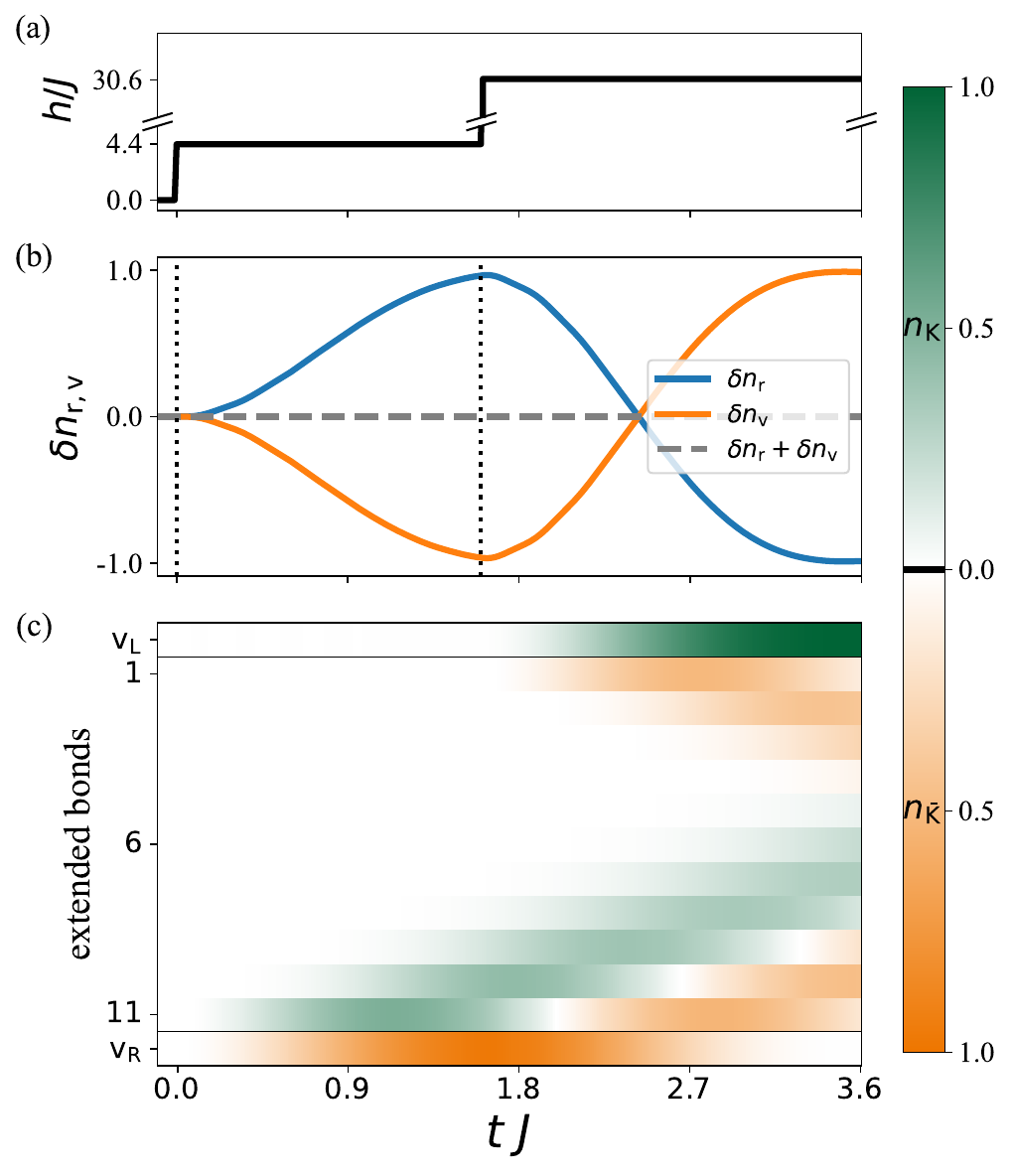}
    \caption{Controlled quench-induced $\text{K}-\bar{\text{K}}$ asymmetry. (a) The quench sequence of $h$. 
    (b) The $\text{K}-\bar{\text{K}}$ asymmetry for real- (blue) and virtual bonds (orange).
    (c) Quench induced time evolution of the quasiparticle numbers $n_{\text{K}}$ and $n_{\bar{\text{K}}}$,
    with the indices $\text{v}_\text{L}$ and $\text{v}_\text{R}$ referring to the virtual bonds and indices $i\in[1,11]$ denoting the real bonds.
    }
    \label{fig:3}
\end{figure}

Figure \ref{fig:3} presents the result of a bi-step quench protocol with the initial state $|\widetilde{\text{vac}}\rangle$, for which the calculation is done in the sublattice-kink basis 
with $\hat{H}_{\text{kink}}^{\text{ext}}$. 
The bi-quench protocol of the boundary tilt potential $h$, is shown in Fig. \ref{fig:3}(a).
The first and second quenches of $h$ set to the condition of $\mu_\text{K}=0$ and $\mu_{\bar{\text{K}}}=0$, 
and resonantly couple the single-$\text{K}$ manifold to SK-vacuum states and the manifold with 
$(n_{\text{K}}=1,n_{\bar{\text{K}}}=0,2)$, respectively. 
Figure \ref{fig:3}(b) quantifies the $\text{K}-\bar{\text{K}}$ asymmetry in the bi-quench process
via $\delta n_{\text{r}\left(\text{v}\right)}$, i.e. the number difference between $\text{K}$ and $\bar{\text{K}}$ 
in the real (virtual) bonds. The dynamics starts from the symmetric population with $\delta n_{\text{r}\left(\text{v}\right)}=0$, 
and upon the first quench, the $\text{K}-\bar{\text{K}}$ asymmetry arises with $\delta n_{r}$ growing to positive values,
indicating that the kink-number conservation is explicitly broken in the real bonds with a favor of $\text{K}$ production.
Applying the second quench, $\delta n_\text{r}$ decreases to negative values,
and the system switches to the $\bar{\text{K}}$-favored situation.
This demonstrates that the $\text{K}-\bar{\text{K}}$ asymmetry cannot only be generated during the quench dynamics,
but also well controlled. Moreover, the figure also shows that, during the bi-quench process,
the $\text{K}-\bar{\text{K}}$ asymmetry for the real bonds is compensated by that in the virtual bonds, 
with $\delta n_\text{r}+\delta n_\text{v}=0$ in the extended lattice.

Figure \ref{fig:3}(c) further visualizes the real-virtual coupling channels in the bi-quench dynamics,
by the density evolution of $\text{K}$ and $\bar{\text{K}}$ in the extended lattice. 
The figure illustrates that, the first quench generates the real-virtual pair excitation
at the right end, and the excited $\text{K}$ propagates into the bulk of the real bonds,
with the $\bar{\text{K}}$ fixed to $\text{v}_\text{R}$.
The second quench not only releases the $\bar{\text{K}}$ on $\text{v}_\text{R}$ to hop into
the real bonds but also excites another real-virtual pair excitation channel, 
adding one more $\bar{\text{K}}$ in the real bonds.
In Fig. \ref{fig:3}(c), it can then be found that both the real-virtual coupling channels can
be selectively activated for the $\text{K}-\bar{\text{K}}$ asymmetry in the quench dynamics.

\textit{Summary and discussion.} This work demonstrates the emergence of intrinsic charge-conjugation violation 
in a one-dimensional bipartite fermionic lattice system, 
in which the SK exhibits measurable population asymmetry in both eigenstates and quench dynamics. 
The experimental realization of this system is very well feasible within current ultracold atomic platforms,
of which the essential building blocks, namely the bipartite lattice structure, the non-local interactions, 
and the local tilt potentials, have all been well developed with ultracold atomic ensembles. 
The bipartite lattice can be implemented using double-well superlattices \cite{superlattice_PhysRevLett.107.255301, superlattice_Science.319.295, superlattice_Nature.448.1029, superlattice_Nature.448.452, superlattice_PhysRevA.73.033605, superlattice_Science.340.1307, superlattice_Nature.587.392, superlattice_Science.377.311, superlattice_PhysRevRes.5.023010, superlattice_PhysRevLett.135.101902, superlattice_PhysRevLett.134.053402, superlattice_Nature.642.909}, a well-established configuration 
extensively employed in quantum simulation and quantum computing studies. 
The non-local interactions are achievable through dipolar atoms or Rydberg-dressed fermionic atoms \cite{non-local_interaction_Rydberg_dressing_fermion_PhysRevX.11.021036} and 
also the hard-core bosonic atoms \cite{non-local_interaction_Rydberg_dressing_boson_Science.390.849}, which can both guarantee single-site occupancy constraints. 
Local tilt potentials can be precisely engineered via site-resolved 
addressing techniques using focused laser beams \cite{addressing_Nature.471.319, addressing_Nature.572.358, addressing_PhysRevLett.114.100503, addressing_Science.352.1562, addressing_PhysRevLett.115.043003, addressing_Nature.621.734, addressing_Science.377.885}. All these building blocks represent mature technologies and can be assembled on the same platform.

Our investigation has been primarily focused on the $\text{K}-\bar{\text{K}}$ asymmetry under the quench dynamics,
and it is desirable to investigate the effect of CCV on other dynamical processes, such as the quasiparticle decay and conversion process.
This can be done by releasing the CSO truncation of the Hilbert space and take into account the effects of
the double and null occupations of a given cell, i.e. the doublon and holon defects, respectively.
The doublon and holon can be viewed as a (quasi)particle-antiparticle pair, 
which can undergo the conversion to  a $\text{K}-\bar{\text{K}}$ pair through the inter-cell hopping during the holon-doublon collision.
Moreover, the doublon and holon serve as boundaries for CSO segments, and can seed the kink-number violation processes. 
The extension then allows the generalization of the CCV test to the particle decay scenarios, 
where doublon-holon annihilation processes can excite imbalanced $\text{K}$ and $\bar{\text{K}}$ 
excitations, providing a more comprehensive platform for studying symmetry-breaking phenomena in correlated quantum systems.

\textit{Data availability.}
The data that support the findings of this letter are publicly available in Zenodo \cite{dataset}.

\begin{acknowledgments}
This work was supported by the Key Researh and Development Program of China 
(Grants No.2022YFA1404102, No. 2022YFC3003802 and No. 2021YFB3900204 ).
\end{acknowledgments}

\bibliographystyle{apsrev4-2_shortauthors}

\makeatletter
\let\APSorigHref\href
\let\APSorigUrl\url
\let\APSorigEprint\Eprint

\renewcommand{\href}[2]{#2}
\renewcommand{\url}[1]{#1}

\providecommand{\Eprint}[2]{#2}
\renewcommand{\Eprint}[2]{#2}

\bibliography{refs}

\let\href\APSorigHref
\let\url\APSorigUrl
\let\Eprint\APSorigEprint
\makeatother

\onecolumngrid
\par\addvspace{3\baselineskip}
\twocolumngrid

\beginendmatter
\textit{Charge conjugation transformation.}
The $\text{K}-\bar{\text{K}}$ exchange is performed by the charge-conjugation transformation $\hat C$, as
\begin{equation}
    \hat C\hat\phi_j^{(\dagger)}\hat C^{-1} = 
    \begin{cases}
        \hat{\bar\phi}_j^{(\dagger)}, j\in\qty[1, N-1] \\
        \hat\phi_j^{(\dagger)}, \text{otherwise}
    \end{cases}.
\end{equation}
Note that $\hat C$ only exchanges $\text{K}$ and $\bar{\text{K}}$ for real bonds,
while acting as identity for virtual bonds.
As illustrated in Fig. \ref{fig:1}(c) with the 3-cell bipartite fermionic lattice as an example,
in which the basis states under the $\text{K}-\bar{\text{K}}$ exchange symmetry are connected by 
yellow arrows, $\hat C$ is equivalent to the particle-hole transformation for fermionic lattice gases
\begin{subequations}\label{pht}
    \begin{align}
        \hat T_\text{ph}\ket{\text{vac}}_\text{atom} &= \prod_{i=1}^N\qty(\hat a_{i,\text{L}}^\dagger\hat a_{i,\text{R}}^\dagger)\ket{\text{vac}}_\text{atom}, \\
        \hat T_\text{ph}\hat a_{i,\sigma_i}\hat T_\text{ph} &= \text{sgn}(\sigma_i)\hat a_{i,\bar\sigma_i}^\dagger,
    \end{align}
\end{subequations}
where $\text{sgn}(\sigma) = -1$ (1) corresponds to the L (R) site, and $\bar\sigma_i$ denoting the distinct site from $\sigma_i$.

In contrast to the fact that particle-hole transformations are normally antiunitary,
which arises due to the fermionic statistics,
$\hat T_\text{ph}$ is unitary in the bipartite fermionic lattice at half filling. 
This is simply due to the fact that in our setup fermions are separated in different cells,
which can also be replaced by hard-core bosons. The fermionic statistics plays little role and the particle-hole
transformation becomes unitary. Nevertheless, the equivalence between $\hat C$ for SK and $\hat T_\text{ph}$
of fermions still suggests a potential connection between the symmetry of particles and quasiparticles in strongly correlated systems.

\textit{Real-virtual bonds extension.}
A complete description of SKs can be obtained by transferring $\hat{H}_\text{FH}$ to 
the kink picture, expressed by the creation and annihilation operators of SKs.
A prerequisite is to specify, to which SK-vacuum state, 
$\ket{\text{vac:L}}_\text{kink}$ or $\ket{\text{vac:R}}_\text{kink}$, 
the creation operators are applied.
This specification can be realized by 
attaching a virtual left-site-occupied cell to each end of the real lattice \cite{superlattice_PhysRevLett.135.101902}, 
as shown by the dotted gray lines in the top panel of Fig. \ref{fig:1}(a).
This extension of the lattice introduces a left and a right virtual bond for SK, 
labeled as vL and vR bonds, respectively, which differentiates the two SK-vacuum states:
$\ket{\text{vac:L}}_\text{kink}$ remains a vacuum for SK in the real-virtual extended lattice and is denoted as $|\widetilde{\text{vac}}\rangle$,
as sketched in Fig. \ref{fig:1}(b1), while $\ket{\text{vac:R}}_\text{kink}$ becomes an excitation state
of two SKs on the vL and vR bonds (Fig. \ref{fig:1}(b2)). 
In the extended lattice, the SK creation operators are then defined as applied to both the real and
virtual bonds, with respect to $|\widetilde{\text{vac}}\rangle$.

\textit{Analyical expression of SK low-energy eigenstates.}
\begin{figure}[tbp]
    \centering
    \includegraphics[width=0.95\linewidth, trim=40 260 40 240, clip]{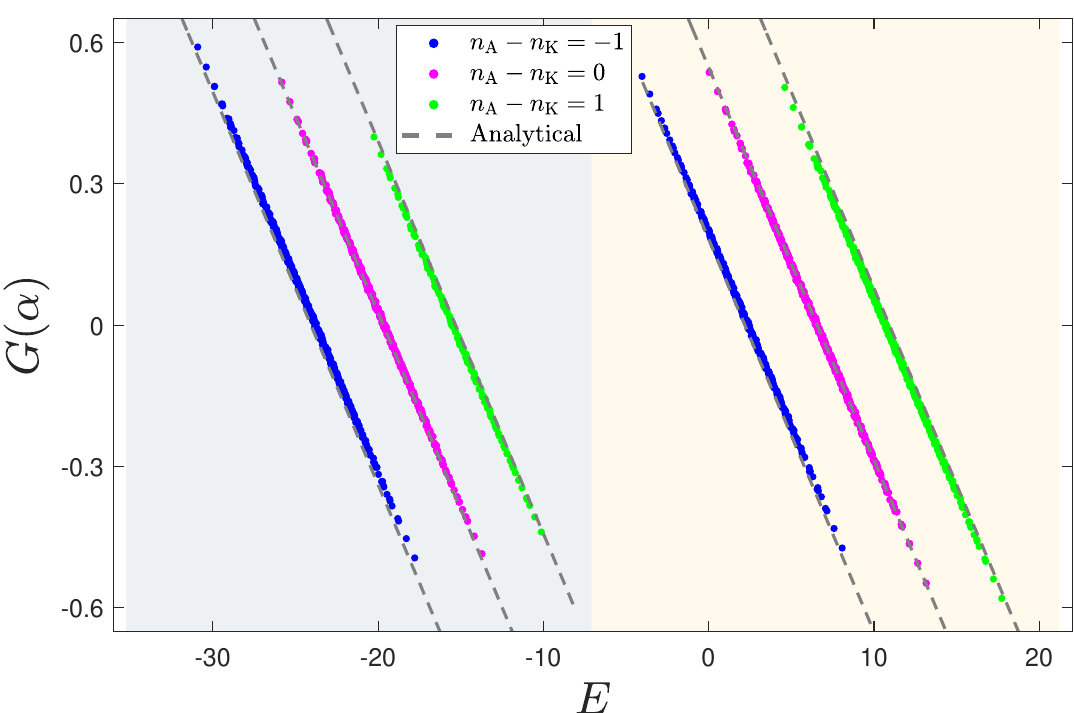}
    \caption{Linearity between $E(\alpha)$ and $G(\alpha)$. Eigenstates of the third- and fourth bands are illustrated, with gray dashed lines being analytical results.}
    \label{fig:EM1}
\end{figure}

Analytical expressions for the low-lying eigenstates can be derived from the conditions that,
(i) SKs are subject to the on-site hard-core interaction,
and (ii), the underlying sublattice geometry mandates a strict spatial alternation between $\text{K}$ and $\bar{\text{K}}$,
forbidding consecutive like-kinks. Taking the eigenstates of $\left(n_{\text{K}}, n_{\bar{\text{K}}}=n_{\text{K}}-1\right)$ for example, 
these conditions collectively define ordered kink basis states, as
$\{ |\vec x=\left(x_1 < x_2 < \dots < x_{2n_{\text{K}}-1}\right)\rangle\}$, 
where $x_{i}$ with $i$ being odd (even) denotes the bond location of a $\text{K}$ ($\bar{\text{K}}$), 
satisfying the alternating kink-antikink pattern.
This constrained configuration maps exactly onto a system of non-interacting spinless fermions. 
Under the condition that the pair production and annihilation are practically suppressed by the chemical
potential of SKs, the mapping gives rise to the expression of the eigenstates as:
\begin{equation}
|\vec{k}\rangle = \mathcal{P} |\vec{k}=\left(k_1 < k_2 < \dots < k_{n_{\text{K}} + n_{\bar{\text{K}}}}\right)\rangle_\text{F},
\end{equation}
where $k_i$ represents a quasi-momentum of a single fermion in an open-boundary $N$-site lattice,
$|\cdots\rangle_\text{F}$ denotes a fermionic Slater determinant eigenstate, 
and $\mathcal{P}$ projects $|\vec{k}\rangle_F$ to the physical kink configuration subspace.
The analytical expression of e.g. the $\alpha$-eigenstate,
leads perturbatively to the linear relationship between the total energy 
$E\left(\alpha\right)$ and the correlation $G\left(\alpha\right)$:
\begin{equation} \label{E-G}
E \left(\alpha\right)= -J G\left(\alpha\right) + \mu_\alpha(n_{\text{K}}, n_{\bar{\text{K}}}),
\end{equation}
where \(\mu_\alpha(n_{\text{K}}, n_{\bar{\text{K}}})=n_\text{K}\left(\alpha\right)\mu_\text{K}+n_{\bar{\text{K}}}\left(\alpha\right)\mu_{\bar{\text{K}}}\) 
represents the total kink excitation energy. 
Numerical validation of this analytical result is presented in Fig. \ref{fig:EM1}, 
which shows excellent agreement between the predicted linear relationship 
and the exact diagonalization results across different subspaces.

\textit{Reshaping of the leaf-like structure and resonance scanning.}
\begin{figure}[tbp]
    \centering
    \includegraphics[width=0.95\linewidth, trim=40 100 20 90]{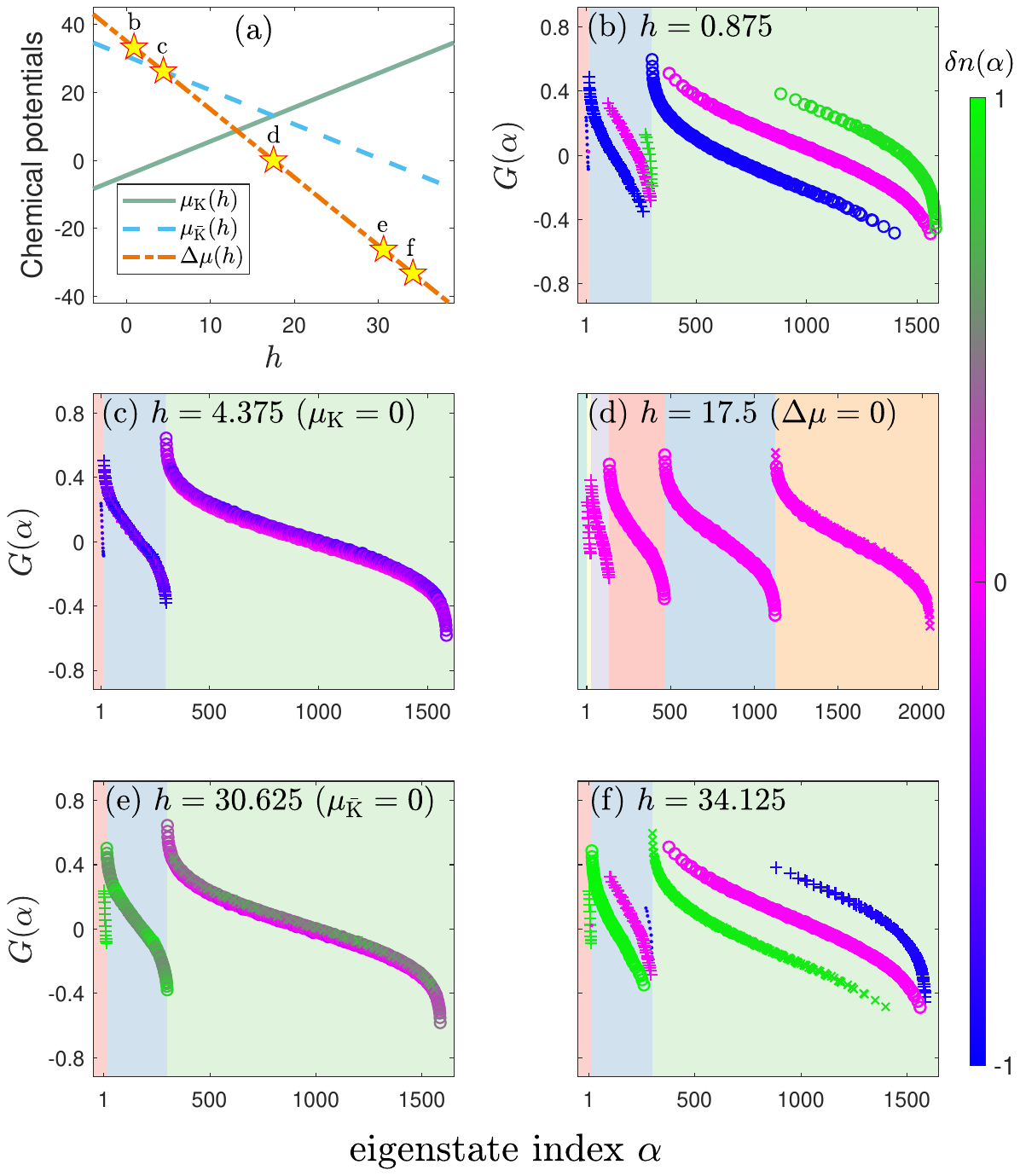}
    \caption{(a) The chemical potential of $\mu_\text{K}$ (green solid), $\mu_{\bar{\text{K}}}$ (blue dashed)
    and $\Delta\mu$ (yellow dotted) as a function of the edge tilt strength $h$.
    (b-f) The leaf-like structures of representative parameters as marked with stars in (a), and
    the choice of color and marker types is the same as in Fig. \ref{fig:2}.}
    \label{fig:EM2}
\end{figure}

The tunability of $\mu_{\text{K},\bar{\text{K}}}$ by the boundary tilts can be illustrated
by the reshaping of the leaf-like structure of the eigenenergy spectrum, 
as presented in Fig. \ref{fig:EM2}(b-f), 
It can be seen that, as $h$ varies, the eigenstates of the same quantum
numbers $(n_{\text{K}}, n_{\bar{\text{K}}})$ remain in the same branch, and different branches undergo a 
lateral motion in the spectrum, with respect to the eigenstate indices. 
Under both weak and strong tilts, as shown in Fig. \ref{fig:EM2}(b) and (f), respectively,
each band in the spectrum preserves the leaf-like structure, 
while the decomposition of each leaf is changed from eigenstates of the same $n_{\bar{\text{K}}}$ to those 
of the same $n_{\text{K}}$, since the band structure is dominated by $\mu_{\bar{\text{K}}}$ and $\mu_\text{K}$ in the
weak and strong regime, respectively.
In the intermediate regime of $h$, the leaves can accidentally fold up,
as shown in Figs. \ref{fig:EM2}(c-e), which correspond to the resonant condition of
$\mu_\text{K}=0$, $\delta \mu=0$, and $\mu_{\bar{\text{K}}}=0$, respectively.

\onecolumngrid
\end{document}